

\magnification= \magstep1
\tolerance=1600
\parskip=0pt
\baselineskip= 6 true mm

\input epsf

\font\smallrm=cmr8
\font\smallit=cmti8

\def\a{\alpha}

\def\d{\delta} 
\def\e{\varepsilon}

\def\k{\kappa}
 
\def\m{\mu}
\def\f{\phi} 
\def\n{\nu}
\def\j{\psi} 
\def\r{\rho}
 
\def\t{\tau}

 \def\W{\Omega}
\def\v{\varphi}

\def\om{\omega}

\def\ddt{{{\rm d}\over{\rm d}t}}
\def\o{\over}
\def\Hin{{\rm {\cal H}_{in}^{BH}}}
\def\Hout{{\rm {\cal H}_{out}^{WH}}}

\def\today{September 27, 1993}

\def\thedate#1{\rightline{#1}}

{\nopagenumbers

\rightline{THU-93/20; UF-RAP-93-11}
\vskip0.05truein{\rm \thedate{\today}\vskip0.4truein}

\vskip 2 truecm

\font\bigg=cmbx10 at 17.3 truept

\centerline{\bigg BLACK HOLE EVAPORATION}
\vskip 0.3truecm
\centerline{\bigg WITHOUT INFORMATION LOSS}
\bigskip
\vskip .5truecm
\centerline{by}
\bigskip
\centerline{C. R. Stephens\footnote{*}%
{\smallrm Email Address: stephens@ruunte.fys.ruu.nl}
, G. 't Hooft\footnote{\dag}%
{\smallrm Email Address: thooft@ruuntk.fys.ruu.nl}
 and B. F.  Whiting\footnote{\ddag}%
{\smallrm Email Address: bernard@bunyip.phys.ufl.edu.  Permanent
Address:  Department of Physics, 215 Williamson Hall, University of
Florida, Gainesville, Fl 32611-8440, USA.} } \bigskip\bigskip
\centerline{Institute for Theoretical Physics}

\centerline{University of Utrecht, Postbox 80 006,}

\centerline{3508 TA Utrecht, Netherlands} \vskip 0.6truein {\noindent
\bf Abstract:} \narrower{ An approach to black hole quantization is
proposed wherein it is assumed that quantum coherence is preserved.  A
consequence of this is that the Penrose diagram describing
gravitational collapse will show the same topological structure as flat
Minkowski space. After giving our motivations for such a quantization
procedure we formulate the background field approximation, in which
particles are divided into ``hard" particles and ``soft" particles. The
background space-time metric depends both on the in-states and on the
out-states.  We present some model calculations and extensive
discussions. In particular, we show, in the context of a toy model,
that the $S$-matrix describing soft particles in the hard particle
background of a collapsing star is unitary, nevertheless, the spectrum
of particles is shown to be approximately thermal. We also conclude
that there is an important topological constraint on functional
integrals.} \vfill \eject
}
\noindent {\bf 1. INTRODUCTION} \medskip The formalism needed to
describe the relevant degrees of freedom in quantum gravity as well as
the laws according to which these degrees of freedom evolve are at
present very much open to conjecture.  For example, even today, it is
still not known how the process of gravitational collapse should be
described. Despite the considerable amount of enthusiastic research
carried out so far, it has not been possible for a general consensus to
be reached concerning whether quantum coherence will be maintained for
the ``black holes" which form as a result of classical collapse, or
whether it gets lost. The latter prospect is often raised as an
inevitable consequence of the thermal properties of the Hawking
radiation following collapse [1], in conjunction with constraints on
the information content of the wave function for the quantum state of a
matter field outside the collapsing object.

Before proceeding further, a word of explanation about the terminology
we shall be using is in order.  The original notion of a black hole
relied upon a global construction for its definition.  The prospect of
black hole evaporation by Hawking radiation has lead to a certain
practical modification of this notion.  But if quantum effects so alter
the course of gravitational collapse that a real curvature singularity
never actually forms, it is clear that a much greater modification is
required.  Certainly a large heavy object will remain present for a
very long time following stellar collapse, since quantum evaporation of
a black hole considerably larger than the Planck mass proceeds very
slowly. It is therefore both convenient and sensible to continue to
refer to this object as a black hole, even if ultimately there are no
outgoing null rays which fail to reach infinity (i.e. no event horizon
nor any true curvature singularity is formed by the collapse).
Thus, in keeping with common practice, we will often refer to the
outcome of gravitational collapse as the formation of a black hole,
mindful that this is a rather classically defined notion. This
definition is not meant to prejudice the final fate of the collapsed
object, whether it be complete evaporation, by radiation correlated or
uncorrelated with the collapse, or whether it become a shadowy exotic
remnant, possessive of bizarre and as yet untold physical properties.
In particular, then, our use of ``black hole'' will refer to certain
properties of the object as seen from afar, and not to the confirmed
existence of a future event horizon or future singularity.

A black hole much larger than the Planck scale possesses many degrees
of freedom, so many, in fact, that it could be regarded as  a
``macroscopic'' system, just as can a salt crystal or a bucket of
water.  Naturally, the question as to whether such a system is quantum
mechanically ``pure", or in a mixed state, will be impossible to decide
on experimentally.  Although the answer would not seem to be important
for classical general relativity, we will argue that the quantum states
of a macroscopic system really matter in the case of gravitational
collapse. In consequence, in this paper we will assume that a black
hole preserves quantum coherence. We emphasize that this is an
assumption, implying that, at present, there is no unassailable
argument to support it, any more than there is to the contrary. Much of
our paper will be taken up with providing a suitable framework in which
to consider the possibility  that quantum coherence remains intact, and
in examining its consequences. For now, by way of motivation, let us
attempt to formulate briefly our reasons for this assumption.

Ultimately, what we would like to obtain is a theory for Planck scale
physics which yields, in a natural way, the quantum mechanical
behaviour with which we are familiar on large distance scales. One
obvious way to safeguard quantum mechanics at large distance scales is
to have exact quantum mechanics already at the Planck scale.  Even if
we thought of describing the ``states'' of the system in terms of
density matrices instead of wave functions, the evolution law for the
density matrix must be defined in such a way that the familiar quantum
mechanical behaviour at large distance scales can still be derived
precisely. If pure states could evolve into mixed states, as the
thermal character of Hawking radiation has been taken to indicate,
then, as will be shown shortly, the evolution law apparently required
for the density matrix in that case seems to have several unpalatable
features. In particular, in some related investigations [2], several
authors have shown that, without quantum coherence, it may be
particularly difficult to obtain a theory with the crucial property
that the Hamiltonian selects out a stable ground state.  Although
energy density might become negative at isolated points, it should not
become unboundedly negative, because this would be entirely at variance
with our general experience. If we relaxed this condition there would
exist no such thing as a stable  state of minimum energy to describe as
a ``vacuum''.  As an additional point, in physics on the Planck scale,
gravitational collapse of small but heavy objects should be happening
frequently and unavoidably. If all these mini-black holes connected our
universe to other universes, and if quantum coherence applied only to
these universes all combined, then quantum coherence would be lost for
our observable universe by itself, and hence we would run into
fundamental difficulties understanding its particular emergence in the
quantum mechanical nature we perceive for the physical world.

A particularly vulnerable assumption underlying the usual arguments
that, in quantum gravity, pure states must evolve into mixed states,
concerns the existence of the classical collapse geometry, with
backreaction effects being, in all respects, effectively small.
Instead, with a different set of assumptions, which include the
existence of a scattering matrix, in this paper we will be drawn to
conclude that the backreaction is able to strongly influence the
evolution of a collapsing quantum system. Since examples have been
found in some special models (e.g. two-dimensional dilaton gravity with
matter fields in the $N\rightarrow \infty$ limit [3]) where information
loss was claimed to be unavoidable, we might expect that requiring the
existence of a quantum mechanically pure scattering matrix will put
severe restrictions on what we will be able to regard as good physical
theories.  A consequence of this type was sought to limit the number of
solution manifolds in superstring theory, and it is a consequence which
we view as rather attractive.  We shall give additional discussion to
support it.  Furthermore, by its effect on classical gravitational
collapse, we do expect that the preservation of quantum coherence will
lead to important new physics, perhaps one example of which may be seen
in our discussion below on the problem of baryon number conservation.
In the following, we will refer to the requirement that quantum
coherence be preserved as the {\it $S$-matrix Ansatz} [4].

At first sight it might seem that the question of whether quantum
coherence gets lost has little to do with physics on Planckian energy
scales. The original derivation by Hawking [1], that the expectation
values of all operators as experienced by late observers are described
by mixed quantum states, seemed to be totally independent of Planck
scale details. Yet, the argument did involve the spacetime geometry
arbitrarily ``close" to the classically determined horizon, and
included energies for which the gravitational redshift had become
arbitrarily large. Moreover, the fact that the outgoing particles look
thermal will be affected by any interactions occuring very near
the horizon and, in turn, these might even reconvert apparently mixed
states back into pure states in such a way that an outside observer
could hardly tell the difference, any more easily than he could for a
bucket of water. Models of transitions between pure quantum states near
the horizon can be formulated by postulating a boundary condition for
all fields on a 2-plane a few Planck distances away from the horizon.
Examples are the ``brick wall'' model [5], the ``stretched horizon''
model [6] and the ``bounce'' model discussed later in this paper. At
present, the distance between the horizon and the 2-plane in question
has to be especially chosen by hand, but that external requirement
should not arise in the full quantum theory. Typically, this distance
increases with the number $N$ of matter fields, and in the
$N\rightarrow\infty$ limit one may therefore run into a direct
contradiction with observation; this again implies that one may not be
able to choose the matter fields freely in a completely consistent
physical theory.  In fact, as we have already intimated, such
restrictions are frequently sought, precisely to exercise a
well-behaved control over the nature of particular theories.

How could one turn an apparently mixed state back into a pure state? As
one way to get some feeling for the kind of possibility which may
arise, consider a quantum theory described by a Hamiltonian $H(\a)$,
that depends on a parameter $\a$ (such as the fine structure constant;
the cosmological constant, $\Lambda$, has sometimes also been taken to
be such a parameter). Suppose that at $t=0$ we have a pure quantum
state $\j(0)$. For each $\a$ we have of course a pure state also at
time $t$ $$\j(t)=e^{-iH(\a)t}\j(0)\eqno(1.1)$$ But now suppose that
$\a$ is poorly known; it has a probability distribution $P(\a)$. Then
the best possible prediction we can give for the expectation value for
an operator $\cal O$ is $$\langle {\cal O}\rangle=\sum_\a
P(\a)\langle\j(t,\a)|{\cal O}|\j(t,\a)\rangle = {\rm Tr}\big(\r(t){\cal
O}\big)\eqno(1.2)$$ which is a mixed state. Now, in the gravitational
collapse problem, it is not difficult to point to possible causes for
such a mixing mechanism: a black hole may be made in numerous different
ways, roughly $\exp{A/4}$, where $A$ is the horizon area. When we
experiment with a black hole at sufficiently late times, all the data
characterizing its more distant past may effectively act as a thermal
heat bath.  The slight uncertainty in components of the Hamiltonian
$H$, conveyed by the $\a$-dependence in our example, can easily be
attributed to the fact that each time we experiment with a black hole
we will be dealing with a specimen that is slightly different from the
previous one:  for example, its expected mass will have some
uncertainty.

It seems quite clear, according to present understanding, that any
evolution based on the assumption that a classical collapse geometry
precedes the  evaporation by Hawking radiation is fundamentally
different from an evolution throughout the entirety of which there is
assumed to exist a well defined scattering matrix. Crucial to this
latter assumption is the observation that backreaction effects induced
by the essential quantum nature of collapsing matter must become
important even before an horizon forms, if ultimately a spacetime
singularity, and with it an inescapable loss of quantum coherence, is
to be avoided.  In an attempt to understand better the consequences of
adhering to a classical determination of the geometry throughout
collapse, Hawking has proposed an alternative quantum theory for any
matter fields accommodated within a curved spacetime geometry.  In
connection with a discussion of the time development of the apparently
mixed final state, he suggested that, in terms of the density matrix
$\r(t)$, the evolution law, which in conventional quantum mechanics is
$$\ddt\r(t)=-i[\r(t),H]\eqno(1.3)$$ should be replaced by a more
general linear equation $$\ddt\r(t)=\$\r(t)\eqno(1.4)$$ where $\$ $ is
a general linear operator. Not surprisingly, because of the rigidity of
our existing quantum theory, a number of difficulties have apparently
arisen as consequences of this proposal, and we now elaborate on
additional reasons for the alternative choice we make. In practical
respects, the space of density matrices $\r(t)$ can be viewed as the
direct product of the space of Dirac bra-states $\langle \j_j(t)|$ and
the space of ket-states $|\j_i(t)\rangle$. In pure quantum mechanics
the ket-states evolve according to the usual Schr\"odinger equation
with Hamiltonian $H$; the bra-states, being the complex conjugates,
evolve with $-H$. Taken together, a state $\r_{ij}=\sum |\j_i(t)\rangle
\langle\j_j(t)|$ evolves according to $$\ddt
\r_{ij}(t)=-i\big(H_{ik}\r_{kj}(t)-H_{jl}\r_{il}(t)\big)\eqno(1.5)$$
which of course is exactly eq. (1.3). But we also see that the
bra-states act in all respects as states effectively with negative
energies! As long as the bra-states and the ket-states do not interact
mutually, as in eq. (1.5), there will be separate energy conservation
in the two half spaces. Energy conservation then guarantees that the
vacuum state $|\W\rangle\langle\W|$ is absolutely stable. In the
ket-space the stability is due to the fact that all excitations carry a
higher energy than the vacuum, and in the bra-space it is due to the
fact that all excitations carry energy less than the vacuum state. Thus
the vacuum state has nowhere to go to.  However, as soon as we deviate
from this, as in eq. (1.4), there will be the possibility for some
effective interaction between the bra's and the ket's. Conservation of
total energy now implies that any exchange among the two halves may
raise the energy of a ket-state while lowering the energy of a
bra-state. Total energy is conserved, but the system is kicked out of
its vacuum. One cannot exclude the complete destruction of any
candidate vacuum state, because although energy is conserved overall,
i.e. the energy for the ket's minus the energy for the bra's is
conserved under evolution via eq. (1.4), that would not at all be the
case for the individual states. Moreover, since phase space of the
separately excited states is infinitely larger than that of the vacuum,
there is no hope of a quick return to the vacuum for any system which
has once departed from it.  In addition, it seems highly unlikely that
this disease would not persist at all larger distance scales,
eventually rendering the theory at variance with observation.

In discussions of the final state of evolution following gravitational
collapse, there has appeared another particular question which we now
briefly address:  it concerns the problem of the non-conservation of
global charges, for simplicity often referred to as ``baryon number".
We consider a specific framework for the formation and evaporation of a
given black hole, and suppose that the number of internal states of the
black hole at some time is given by its entropy:  or is, at any rate,
finite. Then, if quantum states evolve from pure states into pure
states, baryon number conservation must be violated as has been argued
by numerous authors, for example [7]. Consequently, either global
symmetries no longer ensure the existence of corresponding conservation
laws, or they must be violated. We should try to understand where this
violation comes from, or more specifically, how this feature could ever
have been obtained in a theory where one starts out with a baryonic
$U(1)$ symmetry.  In the scheme we have already discussed above, there
emerges a simple way to reconcile this problem. One may take the view
that what is computed in Hawking's calculation really corresponds to a
distribution of Hamiltonians $H(\a)$, where $\a$ represents a shorthand
notation for all ``internal" degrees of freedom that are responsible
for the entropy, $A/4$, of the black hole. Every particular black hole
state corresponds to only one particular Hamiltonian in this
distribution. More precisely, overall, we have just one huge
Hamiltonian, but if we ignore the complicated past history of a
particular black hole, only a small segment of $H$ is applicable to it,
and for each specimen a different segment. Now the entire distribution
will be exactly $U(1)$ invariant, but each particular element,
understood in the way we have specified, violates $U(1)$. So if, at a
later stage, we wished to establish precisely which element of the
distribution applied to a given black hole state, we will be forced to
abandon the $U(1)$ symmetry. This will also have to be done when
specific models for ``brick walls'' or ``stretched horizons'' are
constructed. In later sections of this paper we will not encounter any
baryon violation among the ``soft" or quantum particles we consider; it
will be permitted to arise exclusively via the ``hard" particle
transition amplitudes, $\langle\rm out_0|in_0\rangle$, which we
introduce in the next section.  \def\ss{\scriptscriptstyle}

As many readers will no doubt be aware, a number of seriously
formulated objections have been put forward against the idea that
quantum coherence should be preserved for black holes.  Since the
context we wish to establish renders many of these objections harmless,
whereas the context in which they originally arose has a restricted
applicability within our framework, we will postpone a discussion of
these objections to a later section, at which point their limitations
will be more easily seen than is possible at present.  But one point we
will address further here, for immediate clarification.  The question
which we wish to refer to here concerns a possible relationship between
states with support inside the horizon and states which might
characterize the Hawking radiation, entirely outside the horizon.  To
further examine the potential difficulties which such a relationship
would pose, we choose to consider here pure states in a Heisenberg
picture. In this representation of quantum mechanical effects, the
states are time independent but the operators evolve. We concentrate on
the operator corresponding to the energy-momentum distribution
$\hat{T}_{\ss --}({\bf x},t_0)$ at points alongside the future event
horizon. Usually one only considers states for which this operator has
small, certainly finite, values, because these seem to be the only
states one will be able to produce. Next, we discuss the operators
describing any of the features of the outgoing Hawking particles, such
as their number operator, energies, correlations, etc.  Call these
$\hat{a}_H(t_1)$ for short. It is then possible to argue [8] that the
commutator of the operator $\hat{T}_{\ss --}({\rm Horizon})$ on the one
hand, and the annihilation operator, $\hat{a}_H(r,t)$, at a point
labelled by Schwarzschild coordinates $(r,t)$ on the other, grows
exponentially with time, $t$. Therefore, any of the states for which we
had chosen $\hat{T}_{\ss --}({\rm Horizon})$ to be small (such as would
be appropriate for an infalling observer) must have completely
undetermined values for $\hat{a}_H(t)$, if $t\gg 0.$  Correspondingly,
if we wish to make any observation concerning the Hawking radiation at
late times $t$ then the energy-momentum operator, $\hat{T}_{\ss --}$,
in the neighbourhood of the horizon will become excessively large.

In itself, this uncertainty relation would not have been a disaster if
the particles causing the large $\hat{T}_{\ss --}$ had been completely
transparant. But they are not, because they must be associated with a
gravitational field which, because of the infinite energy shifts
involved, has the ability to destroy everything attempting to cross the
horizon, even if that crossing is to take place at different angular
coordinates. Thus, we conclude that one cannot describe Hawking
particles while at the same time one describes observables, i.e.
expectation values of local operators, beyond the horizon. The
corresponding operators have commutators which are far too large. One
must choose the basis in which one wishes to work: either describe
particles beyond the horizon or the particles in the Hawking radiation,
but do not attempt to describe both. Physically this means that one
cannot have ``super observers", observers that register both Hawking
radiation and matter across the horizon. The corresponding operators
have explosive commutators.

Recognition of the large non-vanishing commutators between operators
describing ingoing material and those of outgoing Hawking radiation is
essential in our approach. Those commutators can be seen as the key to
the fact that a description of ingoing matter will require a space-time
metric which, in the ``inside" of the hole, differs vastly from the
metric needed to describe outgoing matter.  As in ordinary quantum
field theories, one may choose as a basis for Hilbert space either the
set of ``in-states", describing all particles that made the black hole,
and ones that may have fallen in later, as they were when they were all
still asymptotically far away;  or alternatively one may choose a basis
for the ``out-states", describing everything that comes out of the
hole, including the constituents of the final explosion. In practice,
these basis elements could be decomposed into numbers of particles with
each particle being represented by a suitably localized wavepacket.
Here, all in-basis elements are those that produce a black hole at a
certain position, mass, etc., and all out-basis elements are those
compatible with this hole with respect to external observers. In all
these cases all inner products, will be of the same order of magnitude,
even though the metrics they would produce inside the hole might be
very different.  In fact, we shall argue that the metric most
appropriate for the out-states alone should always be taken to be that
of a corresponding ``white hole". We arrive at this choice, rather than
something more exotic or something more classical, because a
microscopic quantum theory should possess $PCT$ invariance; even if
this invariance is not directly manifest in the theory, at least the
construction of the theory should not depend on the orientation of
time, and hence on the distinction between black and white holes.
Unlike some authors, we anticipate that the construction of a
microscopic, quantum mechanical theory should be done in a $PCT$
invariant manner, in contrast to the macroscopic, non-quantum
mechanical case. After all, everything we ever put into our theory
(general relativity, quantum field theory) was invariant under time
reversal.

The format of the remainder of the paper will be as follows: in section
2 we give a qualitative overview of our general strategy, introducing
the $S$-matrix Ansatz and showing how, given knowledge of one
$S$-matrix element, one can build up other elements by considering
perturbations about the known one. We also introduce the concept of
``soft" versus ``hard" particles. In section 3 we discuss in some
detail the relevant in and out states which describe gravitational
collapse and black hole evaporation respectively, and discuss further
the notion of soft and hard particle. We argue that a consequence of
our $S$-matrix Ansatz is that the spacetime metric, which the soft
particle $S$-matrix is defined with respect to, is singularity free in
both past and future, modulo a mild conical singularity near the
effective horizon.  After discussing the conical singularity in more
detail in section 4, in section 5 we consider particle production due
to a generic conical singularity. In section 6 we consider a two
dimensional collapse model with a conical singularity showing that it
is possible to get Hawking type radiation and retain unitary evolution.
In section 7 we draw some conclusions and make some further
speculations.

Before embarking on the development we have just outlined, it is as
well at this juncture to make some comment about the intent of this
paper. It is not our aim to present a theory of quantum gravity (we
manifestly do not have one), nor is it to offer yet a definitive answer
concerning whether black hole formation or decay leads inevitably to
the loss of quantum coherence.  It is, rather, to set a course, by the
following of which we believe one can make progress towards resolving
these and related questions. Ultimately the results we produce will in
a large part determine the measure of our success. In the meantime, by
setting out from a clearly defined position, we hope to have a stable
basis from which to evaluate and attain further progress, wherever that
may eventually lead.

\bigskip
\par\noindent{\bf 2. THE STRATEGY}
\medskip
Having briefly outlined some of our motivations in the introduction, in
this section we will discuss more generally the strategy of our
approach, saving more detailed discussion and illustrative calculations
for later sections. The strategy is partially outlined in [9] and is
based on the $S$-matrix Ansatz, for which the formation and evaporation
of a particular black hole configuration can be described completely
by an $S$-matrix element between state vectors in an appropriate Hilbert
space. As mentioned, when we talk about a ``black hole'' we do not
necessarily mean to imply that there exists a true event horizon and/or
a singularity at $r=0$; in fact we will see later that the $S$-matrix
Ansatz implies that the spacetimes one considers should be globally
regular (up to possibly a mild conical singularity).

Obviously our goal is to be able to compute all the elements of the
$S$-matrix, an ambitious task! To make things simpler we first assume
that the inner product ($S$-matrix element) of one out-bra and one
in-ket is given, say $\langle {\rm out}_0|{\rm in}_0\rangle$.  A more
detailed discussion of the in- and out-states will be given in section
3. Our aim will then be to calculate ``neighbouring" $S$-matrix
elements. Because we wish to work here in an $S$-matrix formulation, we
must always assume, throughout the paper, that asymptotically there is
a well defined notion of time associated with inertial observers very
far away from the ``interaction region'', and we shall see that this
assumption has very definite consequences. What we have in mind, then,
is the following: we begin with a many particle state $|{\rm
in}_0\rangle$, at some very early time, which describes completely, for
instance,  the state of a collapsing star and possible additional
particles. Analogously, the state $|{\rm out}\rangle$ will be taken to
give a complete description of a possible many particle decay mode of
the black hole. We next perturb the in-state
to $|{\rm in}_0+\d_{{\rm in}}\rangle$, or the out-state into  $|{\rm
out}_0+\d_{{\rm out}}\rangle$, or both, where $\d_{{\rm in}}$ and
$\d_{\rm out}$ are
considered to be ``small'' perturbations, and try to calculate
transition amplitudes to perturbed black hole decay modes based on a
knowledge of the amplitude $\langle {\rm out}_0|{\rm in}_0\rangle$.

Two particular questions obviously spring to mind. Firstly, lacking a
full theory of quantum gravity where we could take into account the
metric in a consistent quantum mechanical manner, what background
metric should we use to calculate the matrix elements; and secondly,
what do we mean by a ``small'' perturbation. As far as the first
question is concerned most authors already have a preferred answer.
They usually take the metric associated with the spacetime depicted in
Fig.~1.  This entire spacetime, including the inside of the black hole
(and here an actual horizon and singularity have formed) is, in our
picture, appropriate for describing the in-states.  However, our
out-states cannot be seen as an appropriate unitary evolution of
in-states on this spacetime, due to the lack of completeness of $\cal
I^+$ as a Cauchy surface,  hence this metric violates our $S$-matrix
Ansatz. Fig.~1 results from the assertion that at the horizon the
energy momentum $T_{\ss --}$ should be small. As was already explained in
the introduction, we believe this to be true only for very particular
states, such as the Unruh vacuum, or our in-states, for which there is
a cancellation between the stress energy of the Hawking particles and
the vacuum polarization near the horizon. We do not expect this
cancellation to occur for any of our out-states --- this is not to
say that the entire final state would fail to reproduce this
cancellation, but rather that our basis elements of out-states
do not share this property.
\midinsert
\vskip 4.9 truecm
\epsffile[10 10 50 50]{bhfig1.ps}
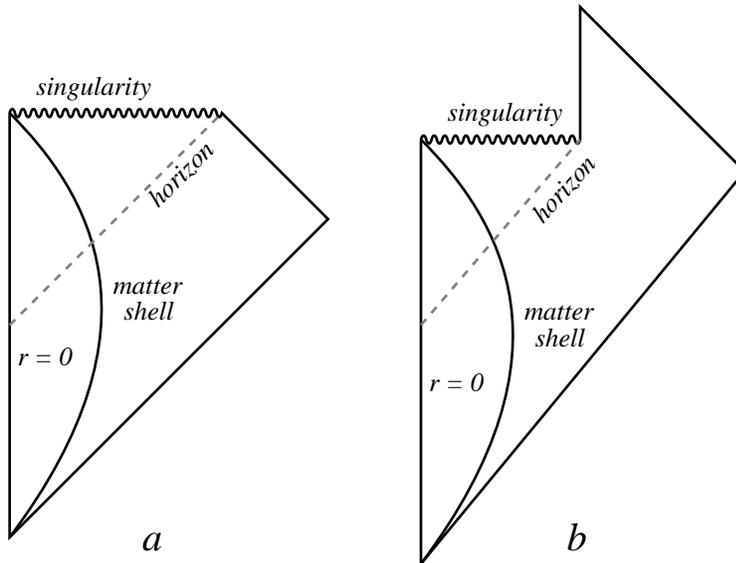
\noindent {Fig.~1. Penrose diagrams of classical gravitational collapse:
a) without evaporation, b) with evaporation; in the standard viewpoint.}
\endinsert
Our aim here is
to be able to compare two different decay modes of a black hole, e.g. a
decay mode corresponding to some given Hawking flux and one which
differs by the addition of a single extra particle. If we consider the
out mode with the one extra particle, then, when it is propagated back
in time to near the horizon, the particle will be seen to have a large
perturbing effect on the geometry. But, our out modes
are supposed to correspond to the results of actual measurements. It is
evident then that, for us, the metric corresponding to the evaporation
of a particular black hole is more appropriately given by the time
reversal of a metric corresponding to some mode of black hole
formation. Thus we take a white hole metric as more suitable for a
description of the out-states. We will later argue that the ``gluing''
together of a black hole and white hole metric will essentially yield a
singularity free metric appropriate for the description of our
$S$-matrix elements.

Note that one needs not at all restrict oneself to out-states
containing the usual spectrum of Hawking radiation (which would give
the hole an expected lifetime proportional to $M^3$ ). All other
out-states compatible with the hole, such as outcoming television sets
or astronauts, will have comparable amplitudes, and are therefore
equally interesting to compute. What makes these out-states much less
probable than thermal Hawking radiation is only the fact that the
latter has much more entropy, or, in particle physics terminology, a
much larger phase space. It is the product $\left|{\rm
amplitude}\right|^2\!\!\times\!{\rm (phase~space~volume)}$ which is
maximal for thermal Hawking radiation.

We now consider a perturbed matrix element $\langle {\rm
out}_0+\d_{{\rm out}}|{\rm in}_0+\d_{{\rm in}}\rangle$. The question of
what we consider here to be a small perturbation is intimately linked
to the question of what metric we should use to compute this matrix
element, as opposed to the metric for computing $\langle {\rm
out}_0|{\rm in}_0\rangle$. To answer this we introduce the notion of
``hard'' and ``soft'' particle. The difference between them will be
associated with the question of whether or not we can ignore the
gravitational effects of the particle. For soft particles this will be
possible, for hard not. Obviously the notion of soft versus hard is not
covariant. We will discuss this in more detail in section 3. Far from
the interaction region there will be a physically sensible notion of
soft versus hard. For instance, if we consider the collapse of a dust
shell we will think of it as being composed of hard particles. One
might also be tempted to think of it as composed of $10^{40}$ soft
particles, however, as our intention is to neglect the gravitational
effects of the soft particles this would not be appropriate. Our
perturbation of this state of hard particles, namely $\d_{{\rm in}}$
and $\d_{{\rm out}}$, will be taken to be due to soft particles. It
should be clear then that we are taking the metric to be determined by
the hard particles. The latter thus give the global metric on which we
can consider the scattering of soft particles. Clearly energy will not
be conserved for the soft particles alone. For global energy
conservation, energy would have to be pumped from the hard particle
background into the soft particles. In fact, this is precisely what
happens in the Hawking effect as portrayed in Fig.~1. Understanding of
the complete interplay between the soft and hard particles will require
a full understanding of the back reaction problem. We will not consider
the latter in any quantitative detail in this paper, but sporadic
comments will be made in reference to where it is important.

Clearly, one could ask how we should calculate $S$-matrix elements
which correspond to perturbations of a given matrix element by hard
particles.  Consider for instance $\langle {\rm out}_0|{\rm
in}_0+\d_{{\rm in}}^h\rangle$, $\d_{{\rm in}}^h$ being a hard particle.
The idea here would be to exploit the non-covariant notion of soft
versus hard to go to a frame where $\d_{{\rm in}}^h$ is soft, whereupon
one could calculate it. Of course, the actual numerical value of the
matrix element cannot depend on the frame in which it is computed,
whereas our ability to compute it might very well be frame dependent.
This strategy would not work for a matrix element such as $\langle {\rm
out}_0+\d_{{\rm out}}^h|{\rm in}_0+\d_{{\rm in}}^h\rangle$ as one would
not be able to find a frame where $\d_{{\rm {\rm out}}}^h$ and
$\d_{{\rm in}}^h$ were both soft. Thus, the distinction we have made
between soft and hard particles, though very useful, is artificial, and
leads only to an approximation scheme for our perturbative
calculations. One could imagine trying to use some cutoff energy,
$\Gamma$, to make the distinction between soft and hard. In practice,
one could then think of adopting a renormalization group approach,
wherein one successively integrates ``shells'' of soft particles to
determine the effective hard particle background in which the soft
particles propagate. This would be a possible approach to the back
reaction problem. What is especially intriguing here is the natural
linking between time and energy scale in a black hole background, i.e.
the later the time at which we observe our particles the smaller scales
(higher energies) they must have been probing near the horizon. In
other words the renormalization group equation that would result from
integrating out shells of soft particles could actually be equivalent
to a time evolution equation for the quantum fields in the evolving
black hole background.

Returning now to the question of what metric to use for the calculation
of matrix elements. We consider the ``effective geometry'' [10]
to be defined by $$ \langle {\rm out}|\hat g_{\m\n}(x)|{\rm
in}\rangle=g_{\m\n}(x) \langle {\rm out}|{\rm in}\rangle\eqno
(2.1)$$ where we work in a Heisenberg picture, so that the operator
$\hat g_{\m\n}(x)$ can simply be defined in any coordinate frame. The
in- and out-states here could be composed of both hard and/or soft
particles, however, only the hard particles will be taken to determine
the metric. The metric for all the soft particle perturbations of this
hard particle matrix element will be taken to be the same. Typically
the in- and out-states will be particular black and white hole
configurations respectively. More motivation for the above choice will
be given using a simple {bf analog} problem in section 3.

Purely as an illustrative analogy, one could also consider the simple
example of non-commuting operators in one-dimensional quantum
mechanics. The Hamiltonian $H={p^2\over 2m}+V(x)$ is neither diagonal
in a $p$-``frame" nor in the $x$-``frame". In neither of these frames
does it make much sense to represent $H$ as a c-number. But we can
observe that $$\langle  p|H|x\rangle = h(p,x)\langle
p|x\rangle\eqno (2.2)$$ where $h(p,x)$ is the usual classical
Hamiltonian in terms of $p$ and $x$ (a feature that is very useful for
deriving functional integral expressions). Note that $h(p,x)$ has
little to do with the expectation values of $H$, either in the given $p$
frame or in the $x$ frame. It therefore need not satisfy the usual
equations of motion.

\bigskip
\noindent{\bf 3. THE IN- AND OUT-STATES}
\medskip
In this section we will explain in more detail how the spacetime metric
that we employ is associated with our choice of in- and out-states. In
most studies of black holes the spacetime metric is chosen to be some
expression for $g_{\m\n}(x)$ that is used as a background determining
the partial differential equations for the quantized fields. This then
means that this (background) metric is taken to be a c-number. Now, if
we view the states in Hilbert space, ${\cal H}$, to be defined in the
Heisenberg picture, then it is clear that in reality the metric should
be an operator just like anything else. Consider a black hole of a
given macroscopic mass, angular momentum and charge, that was formed in
a particular way in a particular region of spacetime. For this
configuration the spacetime metric $g_{\m\n}(x)$ is well-specified
in the space-time region where the collapse takes place, so we say
that we have an approximate eigenstate of the metric operator,
$\hat g_{\m\n}(x)$ in that region. However, our state will
not be an eigenstate of the operator $\hat g_{\m\n}({\bf x},t)$ at
all spacetime points $({\bf x},t)$.  The black hole decays, much like
a radio-active nucleus, and the final explosion may take place at
different instants in time, in different ways. Therefore the spacetime
metric operator $\hat g_{\m\n}(x)$ at spacetime points $x^{\mu}$ near
to where the decay takes place will not be diagonalised at
all\footnote{$\,^*$}{\smallrm Here and in the following
``diagonalization" of course refers to diagonalization of the quantum
operator, not the {\smallit metric} $g_{\m\n}.$}. It should be clear
that there exists no state at all in Hilbert space $\cal H$ that is an
eigenstate of $\hat g_{\m\n}$ at all spacetime points, simply because,
in a Heisenberg picture, these functions are non-commuting operators.
In particular, when black holes occur, the commutators $[\hat
g_{\m\n}({\bf x},t), \hat g_{\m\n}({\bf x}',t')]$ will be large.

Nevertheless one would like to use a metric $g_{\m\n}(x)$ as a
background metric so as to compute the properties of a black hole. So,
let us consider first the in-states, $|{\rm in}\rangle.$ They are
defined just as in formal quantum scattering theory: one assumes that
in the infinite past the system under consideration may have been
formed by a number of ingoing particles which are all far separated
from each other, each described by quantum wave packets, usually with
narrowly but not infinitely accurately defined momenta. We now consider
a particular state, defined in such a way that a black hole is formed
in a well-defined manner and, moreover, so that the metric during
formation is (as nearly as possible) diagonalised. In practice this
seems to be reasonable; a collapsing star for instance completes its
classical collapse at a well-defined moment in terms of a well defined
coordinate grid, and since this is a macroscopic event we do not have
to be worried too much about quantum uncertainties at this stage. Since
we look at one particular black hole, formed in one particular way,
what we actually have here is a linear subspace of Hilbert space, $\cal
H,$ which we will call $\Hin$. The letters {\smallrm BH } here stand
for the given black hole. Different black hole geometries {\smallrm BH}
will correspond to different linear subspaces $\Hin$ of $\cal H.$

Next, as in formal $S$-matrix theory, we can define the out-states
$|{\rm out} \rangle$, in a similar way, as a superposition of states with
widely separated outgoing particles in well-defined wave packets. What
will be new in our approach is that these out-states will also be
limited in such a way that they are taken to be in a linear subspace of
Hilbert space called $\Hout.$ This is the space generated by all states
for which the metric of the exploding black hole is unambiguously
defined. Since this is the time-reverse of an ordinary black hole in
formation, we call this a ``white hole" ({\smallrm WH}). Note that this
way of proceeding obviously fully respects $PCT$ invariance.
It should be clear from the above that for any {\smallrm BH} and any
{\smallrm WH} the two subspaces $\Hin$ and $\Hout$ have practically no
points in common, except for the zero element of $\cal H$. But they are
not orthogonal to each other in $\cal H$. In fact, the inner products
$\langle {\rm out|in}\rangle$, with $|\rm in \rangle \in \Hin$ and
$|\rm out \rangle \in \Hout$, are indeed the $S$-matrix elements we are
after.

Next we wish to observe that there are certain operations one can
perform within the space $\Hin$ or within the space $\Hout$. This
we do by making use of our previously introduced distinction between
soft particles and hard particles. Since we will neglect the
gravitational fields of the former we will be able to consider
quantum mechanical superpositions as well as creation and annihilation
of them without directly being concerned about gravitational back
reaction effects. As stated earlier, the assumption, that in our system
there are at least some particles which are sufficiently soft for us to
ignore their gravitational fields, represents sufficient reason to
consider our approach as an approximation rather than an infinitely
precise theory of quantum gravity.  All other particles are hard
particles and they are the ones that are considered to be responsible
for the particular form of the spacetime metric $g_{\m\n}$ either
during formation of the black hole (when we consider  $\Hin$ ), or
during the evaporation (in  $\Hout$ ). Hard particles carry
gravitational fields, but we will have to avoid considering quantum
mechanical superpositions of them. This does not mean that they cannot
be superimposed quantum mechanically, but that such superpositions are not
elements of the spaces  $\Hin$  or $\Hout.$ This is no great disaster
because one may assume that all of $\cal H$ may be obtained by adding
together either all of the spaces $\Hin$, or all of the spaces $\Hout$,
so that all elements of the $S$-matrix can be obtained from these
spaces alone.
Note that, since in general we consider the metrics $g_{\m\n}$ both in
$\Hin$ and in $\Hout$ to be time-dependent, this means that the
eigenmodes of the full Hamiltonian are neither elements of $\Hin$, nor
of $\Hout$. The full Hamiltonian of the world is of course
time-independent, but, in general, the part of the Hamiltonian (-density)
that describes the evolution of the soft particles alone will be
time-dependent (and space-dependent).

Within a Hilbert space such as $\Hin$ one may consider operators such
as a quantum field $\hat{\f}(x).$ However, only if we restrict
ourselves to sufficiently low frequency modes, so that the particles
created by these operators will not affect the metric $g_{\m\n}(x)$ in
any significant way, will these operators act entirely within our
subspaces $\rm {\cal H}_{in}^{BH}.$ Also we must restrict ourselves to
those spacetime points $x^{\mu}$ where the metric was well specified,
in other words where the deviations between the c-number metric
describing the collapsing star and the expectation value of $\hat
g_{\m\n}$ are small. Operators $\hat{\f}(x)$ can be defined both
inside and outside the horizon. What the operators outside the horizon
mean physically is quite unambiguous. But at this stage what they mean
inside the horizon is somewhat more obscure. Of course one could
extrapolate the field backwards in time, since we are dealing with
Heisenberg states, but that does not alter the fact that these
operators seem to have no effect at all on the outgoing particles.

Consider the ingoing and outgoing particles together. Our strategy
will be to consider first one inner product, $\langle {\rm
out_0|in_0}\rangle = A_0$, as given. All particles in the out- and
in-states here may be hard, or some of them may happen to be soft. Now
consider a small change, either in $\langle {\rm out}|$, or in $|\rm
in\rangle$, or in both. In general $$ \langle{\rm out_0 + \d_{out}|in_0
+ \d_{in}}\rangle = A_0 A_1\eqno (3.1)$$ where now it is crucial that
the changes induced in the in- and/or out-states are entirely due to
the addition or removal of soft particles only. Thus we assume that
neither in the in-state considered nor in the out-state the spacetime
metric is appreciably affected by the small change. We now claim that
the effect, $A_1$, these changes have on the amplitude, may be calculated
by following the evolution of the soft particles in a metric $g_{\m\n}$
that is obtained by combining the metric $g_{\m\n}$ as it was
given at the ``early" spacetime points $x^{\mu}$ by the in-state
configuration, with how it is specified by the out-states at the
``late" spacetime points (where the black hole, now a white hole,
evaporates). The combined metric will be obtained by gluing the
two pieces, the one that is well-specified for the in-states and the
one that is well-specified for the out-states, together. Let us
emphasize again that the combined metric $g_{\m\n}$ does not
necessarily satisfy the usual classical
equations of motion, in particular at the
points where future and past horizons meet --- we will discuss this
procedure more thoroughly in the next section. The evolution of the
soft particle(s) under consideration is obtained by doing quantum field
theory on this combined metric. It is important to note that, since we
wish to limit ourselves to the quantum evolution of soft particles, we
must limit ourselves also to sufficiently low frequency modes of these
quantum fields.

Let us define this combined metric as in eq. (2.1), where now
$g_{\m\n}(x)$ is indeed just a c-number. It must represent both the
black hole for the in-states and the white hole for the out-states.
Within the set of inner products represented by eq. (3.1) (actually
$S$-matrix elements) we may still consider all sorts of small
perturbations on the states, $|{\rm in}\rangle$ and $|{\rm out}\rangle$,
as long as their gravitational effects are kept small.  Hence we may
consider all field operators $\hat{\f}(x)$ for $x^{\mu}$ anywhere in
this spacetime. It is these field operators that we use in order to
deduce the $S$-matrix elements $\langle {\rm out_0 + \d_{out}|in_0 +
\d_{in}}\rangle$ once a single amplitude $\langle {\rm out|in}\rangle$
is given.

Let us stress again that our strategy amounts to making an
approximation: we are only allowed to consider ``soft" changes in the
in- and out-states for which we will be able to compare the $S$-matrix
elements. If these changes would actually affect the metric because
their gravitational back reaction should not have been ignored, we
would get less accurate results. Thus, the changes should be kept as
small as possible. However, we can put them in a chain; each time we
introduce a change in an in- or out-state we make the minuscule
correction required in the metric for the respective black or white
hole so as to be able to perform further series of changes. By
induction one should be able to scan the entire $S$-matrix.  All our
operators act exclusively on the soft particles, and in terms of these
soft particles alone the $S$-matrix will be unitary by construction, as
long as the spacetime generated by $g_{\m\n}(x)$ carries no fundamental
singularities.

Clearly, the distinction between soft and hard particles is artificial,
and is only made for the purpose of being able to do calculations. A
particle may be considered ``soft" with respect to one coordinate grid
but ``hard" with respect to another. Also, if we take a large crowd of
soft particles, their combined action may be better viewed as that of a
hard source. It is an essential part of our strategy to make full use
of the fact that for borderline cases, where a particle could be
regarded as being either soft or hard, the physical values of the $S$
matrix amplitudes should not depend on whether a particle is taken to
be soft or hard. So we vary our choices for the states inside $|\rm
in_0 \rangle$ and $|\rm out_0\rangle$ on the one hand and those in
$\d_{\rm in}$ and $\d_{\rm out}$ on the other. One might suspect that
this gives us a large amount of freedom, but in the most interesting
cases the restrictions on the $\d$ states are severe: they must be soft
as seen by distant observers in the in states as well as the out
states. Both the distant observer of the in-states and the distant
observer of the out states live in nearly flat spacetime and they each
have their preferred coordinate frames.  Considering now the fact that
these two sets of coordinate frames in our metric $g_{\m\n}$ will
produce large blue-shifts in all waves connecting the in-world with the
out-world (through the black hole) we may have to limit ourselves to
alarmingly low energies. There are still two possibilities however: one
is to perform Lorentz transformations, with $\gamma$ factors that are
small enough that they cannot turn soft into hard particles, in these
preferred frames for the distant observers. This may enable us to
stretch the use of soft particles somewhat. The other possibility is to
restrict oneself to variations either in the in-state or in the out
state but never in both at once. This way one can also scan the entire
$S$ matrix, but it will be harder then to check whether it will be
unitary.

Energy and momentum are conserved only among the hard particles; this
could not have been otherwise since they are the sources of
gravitational fields. The soft particles are not restricted by
energy-momentum conservation, since they live in a background that
usually has no Killing vector. Of course, they carry only small amounts
of energy and momentum anyway, and they can absorb or deposit what
little energy and momentum they have into the vast quantities of
background material. The background matter on the other hand is
usually chosen with space and time coordinates that are so
well-specified that, by the uncertainty relation alone, its energy and
momentum have a spread large compared to the total energy and momentum
of the soft particles (though still negligible compared to its own
total energy and momentum). Thus the energy and momentum of the hard
particles form a buffer for those of the soft particles. The
amplitudes for this background matter by itself can hardly depend
on the very slight variations in energy and momentum induced by the
(weak) back reactions of the soft particles.

The procedure of choosing a background that depends both on the choice
of the in-state and on that of the out-state has analogues in
``ordinary" physics.  Let us illustrate this by an example from
established quantum field theory. Consider a $\pi^+$ particle decaying
into a $\m^+$ and a neutrino. This is the dominant decay mode, but
because it is associated with a time-dependent electric current,
$J_\m({\bf x},t)$, there are sub-dominant decay modes wherein extra
photons, and/or $e^+\,e^-$ pairs, are emitted. We will treat the pion
and the muon as `hard' particles, and the photons, $e^+$ and $e^-$
particles, as `soft' particles. Now in the real world the current due
to the electrons is as strong as that due to the pion and the muon, so
the distinction `hard' versus `soft' would make little sense. Therefore
consider a world where the electron charge is much smaller than the
pion/muon charge. In such a model the same procedure would work as the
one we are advocating for black holes. In the fields generated by the
time-dependent current $J_\m$, photons and electron pairs are created,
giving higher order corrections to the $S$ matrix. Unlike the case in
gravitational collapse, where we have no accepted theory of quantum
gravity, here we can in principle calculate the full amplitude in a
totally quantum mechanical manner. Suppose for a moment that we were
restricted to a semi-classical analysis. What would be the appropriate
``classical'' current with which to calculate our quantum mechanical
amplitude? Given that we wish to compute the amplitude for a transition
between a given in state and a given out state the answer is $$J_\m(x)=
{\langle {\rm out}_0|J_\m(x)|{\rm in}\rangle\over\langle {\rm
out}_0|{\rm in}\rangle}\ \eqno (3.2)$$ where the subscript {\smallrm 0}
refers to the state with only the muon and the neutrino present.  The
main reason such a classical current would be appropriate is that it
would contain the effects of the created muon also in the current.
Closer examination of eq. (3.2), to be used as a classical `background'
current, is indeed instructive. If, for instance, we restrict ourselves
to in- and out states that are such that they preserve approximately
energy and momentum, then $J_\m(x)$ will be a very smooth function. If,
however, the pion, muon and neutrino are chosen to be in more precisely
localised wave packets, then also $J_\m(x)$ will show a sharply defined
kink. Photons and/or electron pairs emitted by $J_\m$ will then run
away with a certain amount of energy and momentum removed from the
$\pi\,\m\,\n$ system.

\bigskip
\noindent{\bf 4. THE CONICAL SINGULARITY}
\medskip
In this section we would like to discuss in more detail the global
metric we use and to compare it with the metrics standardly used.
The traditional Penrose diagram  for the collapse and evaporation of a
black hole is shown in Fig.~1b. In the spacetime represented by this
diagram there is a singularity formed, which necessarily leads to a
loss of information from incoming quantum states. Furthermore, as the
diagram is usually proposed, the end-point of the evaporation still
contains the singularity, which really must be removed from the system,
so that the resulting geometry cannot be what might be called past
asymptotically complete. In this circumstance it is difficult to see
how to conceive of a consistent notion of outgoing pure states in
such a spacetime.

In the classical collapse geometry, ignoring back reaction, the
bulk of the Hawking radiation is seen not on all of $\cal{I}^+$, but
only asymptotically in the limit of approach to $\iota^+$. This result
has sometimes been used to suggest that, even in the evaporation
spacetime, all Hawking radiation should again emerge along the
``extended'' null cone of the horizon. However, necessarily in the
geometry for this evaporation, there is an ergo-like region outside the
effective horizon, and negative energy flux in across the horizon from
this region will be correlated with outgoing Hawking-like radiation at
$\cal{I}^+$. Thus, the Penrose diagram for this situation serves to
show that, even if the collapse were halted by some as yet unknown
quantum process (a process which, in some sense, we eventually
characterize by our conical singularity), there must still be some
effective Hawking-like radiation.  It is evident, then, that although
singularities necessarily lead to information loss, they are not
necessary for the emergence of Hawking type radiation, section 6
illustrates this point in a rather elegant fashion. Whether correlations
at $\cal{I}^+$ are such as to preserve incoming information still
depends on other details of the scenario. For example, quantum remnants
which somehow remain non-singular may still be able to trap
information, which cannot then be recovered from any correlations
observed at $\cal{I}^+$.

It is clear that singularities imply information loss. So, if we are
ever to understand how information might be preserved, we have to deal
with spacetimes which are (essentially) non-singular. We achieve this
in our approach by having the metric depend both on the in- and on the
out-states. The collapsing and evaporating spacetimes may be put
together, forming one region where the metric has the usual
Schwarzschild form and others where it is nearly flat (see Fig.~2 for
details). The combined metric has no $r=0$ singularity and yet we
obtain a crude model of collapse and evaporation in which an essential
element of the quantum backreaction, namely the mass-loss to
$\cal{I}^+$, is nevertheless explicitly included.  As stated earlier,
the physically most relevant case is a black hole for which the
$\left|{\rm in}\right>$ and $\left|{\rm out}\right>$ states are chosen
to be the ones that are most likely to occur in reality (having the
largest possible values for the product
(amplitude)$^2 \times$ (phase space). In principle there is nothing
against choosing the $\left|{\rm out}\right>$ state to resemble as
much as possible real Hawking radiation, which has an intensity that
increases as $1\over M^2$, but in practice one then finds redshifts and
blueshifts with ratios diverging exponentially as the black hole mass
decreases.

\midinsert
\vskip 13.2 truecm
\epsffile[10 10 50 50]{bhfig2.ps}
\noindent {Fig.~2. Mapping between black hole and white hole to give
topologically trivial background.}
\endinsert

Depending on the coordinate frame chosen, one often finds
that either the region of ingoing matter, or that of the outgoing
matter, or both, are squeezed into narrow shells. This will then
however be a coordinate artifact in that a Lorentz transformation can
always expand out one of  either the in region or the out region at
the expense of the other;  when both are sharply squeezed it means
no more than that a large relative Lorentz boost exists between the
conformal frame chosen and both the in- and out-going frames. All
Hawking particles, including the objects emitted in the final demise of
the black hole, may then appear to be effectively sqeezed into a
single shell of matter. In practice however it will be easier to
consider less likely final states first. Our procedure will be limited
to ``marginal" black holes where the conical singularity is situated
more than a few Planck lengths from the horizon. By so doing we might
be considering a decay mode of the black hole which is unlikely, but at
least computable. In a model that we will study in more detail in
section 6 both the $\left|{\rm in}\right>$ and $\left|{\rm out}\right>$
states, are chosen to be single shells of matter, separated by a time
interval such that the shells do not come closer to where the horizon
would form than a few units of a Planck length. In addition there will
be clouds of soft particles. The outgoing shell in this model will
result from a ``bounce'' of the ingoing shell and can be thought of as
being due to an explosion of unknown origin in the vicinity of the
horizon.

Now let us look at the process of combining the in and the out metrics.
As pointed out in the previous section, we wish to consider metrics
suitable for discussing both $\left|{\rm in}\right>$ and $\left|{\rm
out}\right>$ states, and we might assemble these so that the metric
operator would then be given as in equation (2.1). A suitable metric at
early times is given by the classical metric of a black hole, whilst at
late times, that of a white hole. To match the metrics of these two
spacetimes together, along a spacelike surface $\tau=\tau_{\ss match}$
as shown in Fig.~2, $\tau$ being some suitable time coordinate which is
regular across the horizon, we are forced to throw two parts of
spacetime away. These are shaded in Fig.~2. Note also that the
crosshatched regions, behind the horizon, are actually inside the
star.  These are the parts near the center of the coordinate frames
which are still practically uncurved and as such allow for a direct
identification.  The resulting spacetime is shown on the right in
Fig.~2, where we have strongly boosted the incoming and outgoing matter
{before we made the identification along $\tau_{\ss match}$}.  The
importance of constructing such a global metric is that now quantum
fields $\hat{\f}({\bf x},t)$ can evolve without any apparent
information loss. Not only is our new ``improved'' spacetime free of
the usual Schwarzschild singularity, it also is topologically trivial,
and every point in it has a future far from the black hole so that it
can be observed by distant observers. The fields $\hat{\f}(x)$ then
live in this spacetime. We see that inside both ``horizons" the fields
$\hat{\f}(x)$ continue to operate just as for the in- and out-states
alone, except where the singularity would be formed. There, these new
fields differ from the fields that would live in the singular black
hole metric of Fig.~1. Apparently they, when acting on a state in
$\Hin$, produce a state outside this subspace of Hilbert space. Only
when sandwiched between our in- and out-states do these operators
behave decently.

Where the in- and outgoing spacetime metric are in conflict with each
other we keep the part that would be visible to the outside observer
and dismiss what is behind the horizon. We also keep the parts that can
be glued together nicely, such as the central regions. This leaves only
one small region where the metric is ill-specified. It is where ingoing
and outgoing matter meet, $S$ in Fig.~2. If we extrapolate the
metric we found as smoothly as possible we still find a large, new kind
of curvature precisely at that spot. In the case of single collapsing
and expanding shells this is a single point in the longitudinal
coordinates, where the curvature is Dirac-delta distributed over an
entire two-sphere. It then
corresponds to what can be described as a conical singularity, comparable
to conical singularities describing particles in 2+1 dimensional
gravity, except that these are the Lorentz equivalents of them:
coordinates are locally flat but when they are compared after a journey
along a closed path around the singularity the connection goes via a
Lorentz transformation.  Scattering of quantized fields around such a
singularity has been considered in [9] and will be discussed here in
section 5 for completeness. It is of crucial importance to observe that
the singularity is so mild that no loss of quantum information is
suffered by the evolving states of soft particles in such a metric.
Indeed, if we replace the shells by more smoothly distributed matter
the singularity is smeared into a non-singular (but still highly
curved) metric. Clearly then, the $S$-matrix, in terms of the soft
particles alone, will be unitary.

We now have a geometry on which to work, which is a natural consequence
of our $S$-matrix Ansatz, and it is perhaps helpful to compare our
result with that of Unruh [1] in his original discussion of the
collapse problem. Essentially Unruh argued that the inside of the star
could be replaced by a region of the Kruskal manifold bounded to the
future by the past horizon, and effectively including the past
singularity. By contrast, we have argued that if complete evaporation
can occur, then at sufficiently late times the resulting spacetime must
again resemble Minkowski space or, in fact, its time reverse as it
occurs inside the collapsing shell. And we have chosen to include this
in such a way that no singularity appears or forms anywhere in the
spacetime, by the reverse of Unruh's process.

\bigskip
\line{\bf 5. PARTICLE CREATION BY A CONICAL SINGULARITY\hfil}
\medskip
The $S$-matrix Ansatz, which requires a (topological) singularity free
spacetime, seems to demand the presence of a conical singularity, or at
least a region of high curvature. We now consider the effect a conical
singularity, in the simplified context of flat space, has on a
quantized state in field theory. Since the metric has no timelike
Killing vector there is no conserved energy. If we begin with the
vacuum state at $t=-\infty$ the state at $t=+\infty$ will in general
contain particles. The computation is not hard. Observe that, in
contrast with the familiar calculation of the Hawking-Unruh effect,
there will be no information loss. Later we will be interested in
different initial states, but let us begin with the vacuum.

For simplicity we take the field to be scalar. The local operator
$\v({\bf x},t)$ is given by $$\v({\bf x},t)=\int d^3k {1\over
\sqrt{(16\pi^3k^0)}}\big(a_k {\rm e}^{ikx}+a^\dagger_k{\rm
e}^{-ikx}\big)\eqno(5.1)$$ where $a_k$ and $a^\dagger_k$ are
annihilation and creation operators at given three-momentum ${\bf k}$,
satisfying the usual commutation rules normalized with a Dirac
delta function in {\bf k} space. As usual we define $k^0=\sqrt{({\bf
k}^2+m^2)}$, $kx={\bf k}\cdot{\bf x}-k^0t$.

We take equation (5.1) to hold at time $t<0$, before the singularity
$S$ occurred. At time $t>0$ we take the fields to be $$\v(y)=\int d^3k
{1\over \sqrt{(16\pi^3k^0)}}\big(b_k {\rm e}^{iky}+b^\dagger_k{\rm
e}^{-iky}\big)\eqno(5.2)$$ \noindent where $y$ are Cartesian
space-time coordinates at $t>0$. They are related to the ${\bf
x},t$-coordinates by $$ y=x\quad {\rm if}\quad x_1<0 \quad ,\quad
y=L^{-1}x\quad {\rm if}\quad x_1>0\>,\eqno (5.3)$$ where $L$ is the
Lorentz transformation $$L=\pmatrix{\cosh\f&\sinh\f&0&0\cr
\sinh\f&\cosh\f&0&0\cr 0&0&1&0\cr 0&0&0&1\cr}\eqno(5.4)$$

Using the fact that both sets of modes are complete one can relate the
two sets of annihilation and creation operators via a Bogoliubov
transformation. One finds that
$$b_p=\int {d^3k\over(2\pi)^3}\big( A^+_{pk}
a_k+  A^-_{pk}a^\dagger_k\big)\eqno (5.5)$$ where $A^+_{pk}$ and
$A^-_{pk}$ are the Bogoliubov coefficients. From now on the variables
$p$ and $k$ are only the $x$-components of the momenta, the ones that
transform non-trivially under the Lorentz transformation (5.4). $p^0$
and $k^0$ are the usual time components of the momenta. Also we write
$x=x_1$. Let us furthermore use the shorthand notation $$\cosh\f=c\quad
,\quad \sinh\f=s\eqno(5.6)$$ where $\f$ is the Lorentz boost parameter.
The Bogoliubov coefficients can be computed to be $$ A^\pm =
{1\over4\pi}\sqrt{p^0\over k^0}\int_0^\infty dx\left(\big(1\pm{k^0\over
p^0}\big)\, {\rm e}^{-i(k-p)x}+\big(1\pm {ck^0 - sk \over p^0}\big)\,
{\rm e}^{i(ck -sk^0-p)x} \right)\d^2(\tilde p-\tilde k)\eqno(5.7)$$
where $\tilde p$ and $\tilde k$ are the transverse momentum
components. The integral over $x$ can of course be calculated:
$$A^\pm({\bf p}, {\bf k}) = {1\over 4\pi\sqrt{p^0k^0}}\left({-i(p^0\pm
k^0)\over  k-p-i\e}+{i\big(p^0\pm(ck^0-sk)\big)\over
ck-sk^0-p+i\e}\right)\d^2(\tilde p-\tilde k)\eqno(5.8)$$ Note that
$A^-({\bf p}, {\bf k})\neq0$, hence the Bogoliubov transformation mixes
positive and negative frequencies and therefore there is particle
production.

The number of particles created in a mode $\bf p$ is given by $\langle
b^\dagger({\bf p})b({\bf p})\rangle_0$, where $\langle\,\rangle_0$
corresponds to the vacuum of the annihilation operators $a_k$. It is
found to be $$\eqalign{ \langle b^\dagger({\bf p})b({\bf p})
\rangle_0&=\int {d^3k\over(2\pi)^3}\vert A^-({\bf p},{\bf k})\vert^2=
\cr &={1\over16\pi^2p^0}\int\! {{\rm d}k\over
k^0}\left({(p^0-k^0)(ck-sk^0-p)+(p-k)(p^0-ck^0+sk)\over
(k-p)(ck-sk^0-p)}\right)^2\cr}\eqno (5.9)$$ where in the integral we
must insert $ k^0=\sqrt{k^2+\tilde k^2+m^2}$, and similarly for $p^0$.
Note that in the absence of an IR cutoff, such as a mass, most of the
particles are created in long wavelength modes. In this case the problem
is clearly scale invariant.

The rest is straightforward arithmetic. All integrals can be performed
and the result is $$ \langle b^\dagger({\bf p})b({\bf p})
\rangle_0={1\over 4\pi^2 p^0}\left[{\f\over \tanh
\textstyle{1\over2}\f}-2\right]\eqno(5.10)$$ For small $\f$ the
quantity between square brackets is $$ [\dots
]={\f^2\over6}-{\f^4\over360}+\dots\eqno(5.11)$$ \noindent and if
$\f$ is large then it approaches $$ [\dots]=\vert
\f\vert-2+2\vert\f\vert{\rm e}^{-\vert\f\vert}+\dots\eqno(5.12)$$
Note that the $\bf p$ dependence is ${{\rm d}^3p/ 2p^0 }= {\rm
d}^4p\d(p^2+m^2)$, which is Lorentz invariant. Invariance under
Lorentz transformations in the $x$ direction is not surprising. But the
invariance in the transverse direction is an accident. The Bogoliubov
coefficients $A^\pm$ themselves do not have this latter invariance.
Also, the fact that eq. (5.10) is independent of the sign of $\f$ is an
accident.

The coefficients of eq. (5.8) were computed for given 3-momenta. The
calculations simplify however if we go to lightcone coordinates
instead. The outcome, such as eq. (5.10), of course stays the same.
Note that we, naturally, have a very singular problem here. Effectively
there is no UV cutoff on the particle production, hence, for instance,
the response rate, which here is constant in time, of a particle
detector would diverge. This problem can be overcome by ``smoothing''
out the singularity with a smoothing function $f({\bf x},t)$.

\def\p{\partial}
\def\ra{\rightarrow}
\bigskip
\noindent{\bf 6. ``BOUNCING'' TWO DIMENSIONAL SHELL MODEL}.
\medskip
\def\Rs{R^\ast}
\def\tin{t_{\ss{\rm in}}}
\def\tbn{t_{\ss{\rm bn}}}
\def\rin{R_{\ss{\rm in}}}
\def\rbn{R_{\ss{\rm bn}}}
\def\rsin{\Rs_{\ss{\rm in}}}
\def\rsbn{\Rs_{\ss{\rm bn}}}
\def\uin{u_{\ss{\rm in}}}
\def\ubn{u_{\ss{\rm bn}}}
\def\vin{v_{\ss{\rm in}}}
\def\vbn{v_{\ss{\rm bn}}}
\def\ras{r^\ast}
We turn our attention now to a model that encapsulates most of the
features we have been discussing earlier: hard and soft particles,
conical singularity, globally Minkowskian spacetime etc. Here we
will examine the consequences of the physics discussed in sections 2
and 3 in the context of another toy model. In the seventies
there was much
work done (see [11] and references therein) on simple collapsing
shell models in two dimensions, the
aim being to try to illuminate the essential physics of the Hawking
effect in as simple and uncluttered an environment as possible. The
model consisted of a shell (in two dimensions obviously a point
particle) on a trajectory $R(\t)$, $R$ being the radial coordinate of
the shell and $\t$ being the proper time as measured in the comoving
frame with the shell. In the language used previously the collapsing
shell here for us represents the effects of the ``hard'' particles. We
wish to analyze the production of ``soft'' particles in this ``hard''
particle background. The metric is given by $$ds^2=-dT^2+dr^2
\ \ \ \ \ \ \ \ \ \ \ \ \ \ r<R(\t)\eqno(6.1)$$ $$ds^2=-\left(1-{2M\o
r}\right)dt^2+{1\o\left(1-{2M\o r}\right)}dr^2 \ \ \ \
\ \ \ \ \ \ \ \ r>R(\tau)\eqno(6.2)$$ So, inside the shell we assume
flat space and outside the standard Schwarzschild metric. In
conventional models $R(t)$ would be required to satisfy the classical
equations of motion. In such models the background would develop the
usual singularity. Ours will deviate from that in the sense that we
will admit one point in space-time where the classical equations are
completely violated. This point is where we connect the in-metric with
the out-metric and will be referred to as ``the bounce''. We will denote
by $\tbn$ the time at which this bounce takes place, and $\rbn$ $(>2M)$
the corresponding minimum radius of the shell. By construction
such a space-time is free of any singularity that would absorb
information. We will assume that collapse ``begins'' at $t=\t=0$, so that
$R(t)=\rin$ when $t<0$, $\rin$ being the initial radius of the
star before collapse. The relation between $T$ and $t$
is such that the metric is continuous across the shell. There will be
curvature on the shell.

For simplicity we will quantize a massless scalar field $\f$ on this
spacetime where
$$\f=\sum_{\omega}a_{\om}u_{\om}+a^{\dag}_{\om}u^{\ast}_{\om}\eqno(6.3)$$
Before the onset of collapse we have scalar field modes $$u_{\om}={1\o
(4\pi\om)^{1\o2}}(e^{-i\om\bar v}-e^{-i\om\bar u})\eqno(6.4)$$ $\bar u$ and
$\bar v$ are chosen so that $\f$ vanishes at the coordinate origin.  We
introduce the shifted null coordinates $$U=T-r+\rin
\ \ \ \ \ \ \ \ \ \ \ \ \ \  \ V=T+r-\rin\eqno(6.5)$$
$$u=t-\ras+\rsin \ \ \ \ \ \ \ \ \ \ \ \ v=t+\ras-\rsin\eqno(6.6)$$
where $\ras$ is the standard tortoise coordinate
$\ras=r+2M{\rm ln}|{r\o 2M}-1|$. We require that the modes (6.4)
correspond to vacuum modes in the asymptotic past, i.e. that they are
positive frequency with respect to $\p\o\p t$.  Hence we take $\bar
v=v$, and this defines the ``in'' vacuum along ${\cal I}^-$. We denote
the relations between the coordinates inside and outside the shell
$v=\beta(V)$ and $U=\alpha(u)$. Now, modes $e^{-i\om\bar v}$ are
``reflected'' off the coordinate origin $e^{-i\om\bar v}\ra e^{-i\om\bar
u}$ thus $$\bar u=\beta(U-2\rin)=\beta(\alpha(u)-2\rin)\eqno(6.7)$$
We can now write the metric as $$ds^2=C(\bar u,\bar
v)d\bar u d\bar v$$ $$\ \ \ \ \ \ \ \ \ =-{d\bar u d\bar
v\o\beta'(U-2\rin)\beta'(V)} \ \ \ \ \ \ \ \ r<R(t)\eqno(6.8)$$
$$\ \ \ \ \ \ \ \ \ =-\left(1-{2M\o r}\right){d\bar u d\bar
v\o\beta'(\alpha(u)-2\rin)\alpha'(u)}
\ \ \ \ \ \ \ \ \ \ r>R(t)\eqno(6.9)$$ where $'$ denotes
differentiation with respect to the argument of the function. As shown
in [12], the renormalized stress tensor in the case at hand is
determined solely in terms of the conformal factor $C$.
$$<T_{\mu\nu}>_{\ss{\rm ren}}=\theta_{\mu\nu}-{R\over48\pi}g_{\mu\nu}
\eqno(6.10)$$
where $$\theta_{uu}=-{1\o12\pi}C^{1\o2}\p_u^2C^{-{1\o2}}$$
$$\theta_{vv}=-{1\o12\pi}C^{1\o2}\p_v^2C^{-{1\o2}}$$
$$\theta_{uv}=\theta_{vu}=0$$ For the moment we will just be concerned
with the stress tensor outside the shell, and in particular with the
outgoing radiation at ${\cal I}^+$, described by
$$T_{uu}={1\o12}F_r(e^{6M\o
r})+\alpha'^2F_U(\beta')+F_u(\alpha')\eqno(6.11)$$ where
$F_x(y)={1\o12\pi}y^{1\o2}{\p^2\o\p x^2}y^{-{1\o2}}$.  One of the nice
aspects of all this is that the expression for $<T_{\mu\nu}>$ is exact
for a given trajectory of the shell.

The above formulae are applicable to an arbitrary trajectory. Here
in line with the physics discussed previously we wish to consider what
happens for a trajectory with a bounce. The explicit trajectory we will
examine here is
$$\Rs(t)=\rsbn\ \ \ \ \ \ \ \ \ \ \ \ \ \ \ \ \ \ \ \ t>\tbn$$
$$\Rs(t)=-{1\o\k}{\rm lncosh}\k t+{\rsin}\ \ \ \ \ \ \ \ \ \ \tbn<t<0
\eqno(6.12)$$ $$\Rs(t)={\rsin} \ \ \ \ \ \ \ \ \ \ \ \ \ \ \ \ \ \ \ \
t<0$$
As mentioned $\tbn$ is the time at which the bounce, or in this case
``the stop'', takes place, and $\rbn$ is the corresponding radius of the
shell. We assume that $\rbn$ is more than a few Planck lengths from $2M$
in order that we may consider any individual particles created by the
collapse to be soft. It is worth noting
that the $\ln\cosh t$ trajectory is the solution of one dimensional
Liouville theory. The point
at which the collapse stops or bounces (``explodes'') is a conical
singularity. The late time form of the radiation before the explosion
or stopped collapse is indifferent to it. The only term which
contributes to $<T_{uu}>$ on ${\cal I}^+$ before the collapse or explosion
comes from the last term in (6.11). To evaluate it we need the relation
between $u$ and $U$. To do this we match the distance element along the
shell in the two metrics to get $$\alpha'(u)={C^{1\o2}(1-{2M\o R}\dot
{R^{\ast}}^2)^{1\o2}-C\dot\Rs\o1-\dot\Rs}\eqno(6.13)$$ where
$C=(1-{2M\o R})$. We now consider the asymptotic form of $<T_{uu}>$ on
$I^+$ for $t>>{1\o\k}$, $t<\tbn$. In this limit
$$R(t)\ra2M(1-e^{-{t\o2M}})
\ \ \ \ \ \ \ \ \dot\Rs\ra-(1-2e^{-{t\o2M}})\eqno(6.14)$$
$${\ddot\Rs}\ra-{1\o M}e^{-{t\o2M}}
\ \ \ \ \ \ \ \ \ {\buildrel{...}\over \Rs} \ra{1\o2M^2}e^{-{t\o2M}}
\eqno(6.15)$$
$$C\ra4e^{-{t\o2M}}\ \ \ \ \ \ \ \ \ \dot C\ra-{2\o
M}e^{-{t\o2M}}\ \ \ \ \ \ \ \ \ \ {\ddot C}\ra{1\o
M^2}e^{-{t\o2M}}\eqno(6.16)$$ Substituting these limits into (6.11) gives
$$<T_{uu}>\ra{\k^2\o48\pi}\eqno(6.17)$$ which is the standard Hawking result.

Of course, globally we now have a very different state of affairs to
that of the standard Hawking effect where the shell collapses into a
singularity. The Penrose diagram for this stopped collapse is shown in
Fig.~3.  It is globally Minkowskian, hence there is no loss of
information. The radiation near $F$, however, will be Hawking like,
i.e.  have a Planckian spectrum to a very good approximation, as long
as $\kappa\tbn\gg1$.  Remember that $EF$ can be a very large amount of
retarded time and that close to $F$ the latter is very constricted.  At
$F$ itself there will be a singular burst of radiation due to the
conical singularity, as mentioned in the previous section by smearing
the singularity this burst can be made finite. We conclude therefore
that it is possible to obtain the ``Hawking effect'' without an
intrinsic loss of information.  Naturally, there are many possible
deficiencies of this toy model not least of which is the reasonableness
of stopping the collapse or making the shell bounce. Our attitude here
is the following: the stop or bounce, as we have presented here, is
just another hard particle background. We are examining the soft
particle effects on this background.  In this sense, within the
confines of our $S$-matrix Ansatz, we are just looking at certain
$S$-matrix elements.  In fact, following our discussion of previous
sections, we would argue that a bounce type phenomenon is a necessary
consequence of the $S$-matrix Ansatz itself.
\midinsert
\vskip 10.5 truecm
\epsffile[10 10 50 50]{bhfig3.ps}
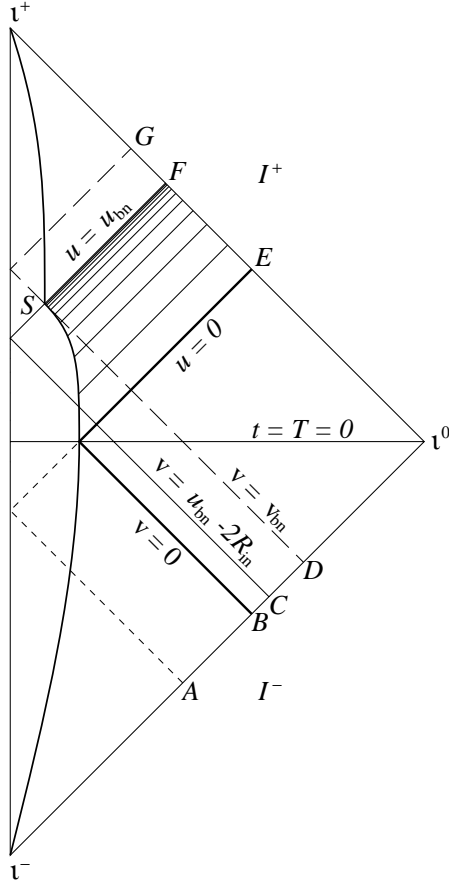
\centerline {Fig.~3. Penrose diagram for the  ``bounce'' model.}
\endinsert
It is clear that the bounce, relative to the standard point of view, is
a radical departure, and can only come about due to large back reaction
effects. The canonical point of view has been that because $\langle
T_{\mu\nu}\rangle$ is small when computed in the Unruh vacuum back
reaction effects cannot be large, and therefore the classical blackhole
geometry should until late times in the evaporation of the hole be a
good description of the geometry. We have presented in previous
sections arguments indicating why we believe the back reaction to be
large. We can now also look at things from a different point of view.
Originally there was an argument as to where the Hawking radiation
originated: should it be associated with the star itself during the
final stages of the collapse, or should it be viewed as something
unrelated to the star, dependent only on the horizon which emerges with
the formation of the black hole. The consensus view eventually became
the latter. In the two dimensional collapse models there was an influx
of radiation $-{\k^2\o48\pi}$, constant in advanced time (neglecting
back reaction), along the entire future horizon which just accounted
for the Hawking flux ${\k^2\o48\pi}$ at ${\cal I}^+$.  However, by
adding in the bounce one can see things in a different light.
Asymptotically there is a Hawking flux. This has unambiguously nothing
to do with the existence of a future horizon as there isn't one! What
one also sees then is the following: in the bounce model, retarded
time, near the bounce time, is very constricted, i.e. the amount of
retarded time which has elapsed between $E$ and $F$ is very large for
bounces which take place close to $2M$. The amount of advanced time,
however, is very short, between $B$ and $D$. In this model then the
back reaction due to the effect that the shell must unambiguously have
lost mass equal to $\int^{\ubn}_{\uin}T_{uu}du$ between $E$ and $F$
must be very large, as all this radiation originates in a region
spanning a very short amount of advanced time, $\vbn-\vin$.  Thus, as
far as the back reaction is concerned, a very different picture arises
here.

We would like, eventually of course, to be able to view the bounce as a
natural consequence of the evolution of a collapsing star as opposed to
a boundary condition which is a natural consequence of our $S$-matrix
Ansatz. This seems to be a distinct possibility. The bounce can be seen
as a device in this particular model which demonstrates that the
Hawking radiation cannot originate in a continous process, constant in
advanced time, in the vicinity of the future event horizon because such
does not exist. Given that this is so one can ask, what happens if we
remove the bounce? The fact that the shell is radiating has nothing to
do with whether it bounces or not. However, one should now explicitly
try to take the backreaction into account. The mass of the shell
$M(\tau)$ then changes with time. Here we are at a disadvantage, of
course, as we have no Einstein gravity in two dimensions. One could
take several options: take the physics of the bounce and examine it in
the context of a ``real'' theory of two dimensional gravity; take the
physics of the bounce and examine it in the context of a four
dimensional collapse scenario; or try and force the back reaction into
the model as it is.  Obviously the second option is the most
preferable, but probably the most difficult. The first brings up the
question of motivation in going to a two dimensional model. The
motivation in the original shell models was essentially just finding a
simple model that mimicked as closely as possible the Hawking four
dimensional calculation. One was not dealing with a ``proper'' two
dimensional theory of gravity but trying to illuminate the essential
features of the four dimensional calculation in a simpler setting. The
collapse trajectories chosen were such that they had the generic form
$R^{\ast}=-t+A-Be^{-\k t}$ which is the generic form of a freefall
trajectory in four dimensions. To take into account backreaction one
might propose an ansatz in the two dimensional case
$R^{\ast}=-{1\over\kappa(t)}\ln\cosh\kappa(t)t+\rsin$ where $\k(t)$
is to be determined merely from
energy conservation. This would be tantamount to assuming that the
object is falling on a geodesic trajectory associated with a
Schwarzschild metric of time dependent mass and could be thought
of as being a sort of WKB ansatz.

There are also other questions to be asked such as where do the
correlations that ensure that there is no intrinsic information loss
exist. To answer this we must consider the full Bogoliubov
transformation. If one propagates modes from ${\cal I}^-$ through
the collapse and bounce then back to ${\cal I}^+$
one finds they take the form
$$u_{\om}={1\o{(4\pi\om)^{1\o2}}}(e^{-i\om
v}-e^{-i\om\beta(\alpha(u)-2\rin)}) \eqno(6.18)$$ An incoming wave
$e^{-i\om v}$ on striking the center of the star bounces out again
becoming an outgoing wave $e^{-i\om\beta(\alpha(u)-2\rin)}$, which
leaves the collapsing star and becomes a complicated function on ${\cal
I}^+$. To get the Bogoliubov coefficients one needs to Fourier analyze
these complicated modes on ${\cal I}^+$, or alternatively, as the
currents are conserved, on any Cauchy surface. Modes that pass through
before the collapse begins, i.e. that begin along
$A\iota^-$ in Fig.~3, are
not affected since the red and blue shifting as the mode goes into the
star and then back out compensate each other. During the collapse there
is an escalating redshift for modes, originating along $AD$,
propagated through the star and out to ${\cal I}^+$. It is the
generic exponential late time form of this redshifting that is
really responsible for the Hawking effect. The bounce occurs, then some
modes come out after the bounce along $FG$ which have been disturbed
during their passage through the star. If one analyzes the modes at
late times $\ubn>u\gg0$ then one finds that the number of particles in
a mode $\om$ is $$n_{\om}={1\o(e^{{2\pi\om\o\k}}-1)}\eqno(6.19)$$ In
deriving this we have assumed that one is interested in frequencies
$\om\gg{1\o \ubn}$. For frequencies that do not satisfy this criterion
the modes are not thermally distributed. When we use the term thermally
distributed here, of course, we mean it in the sense that the spectrum
(6.19) is Planckian. Naturally, the state is a pure one as the
spacetime is topologically Minkowskian.  All the correlations in this case
lie in very low frequenecy modes, which one can think of as
implying correlations between early time and late time radiation. We
will return to these matters in much more detail in a future
publication.

We believe this model to be quite suggestive regarding the situation
to be found in four dimensions. Naturally definitive answers can only be
given after the full back reaction problem is solved in four
dimensions. In particular one might well expect that the (smeared)
conical singularity we have exploited here would turn out to be a
natural consequence of stellar collapse being then a representation of
the final black hole explosion.

\bigskip
\line{\bf 7. DISCUSSION AND CONCLUSIONS \hfil}
\medskip
As we lack a full theory of quantum gravity, one of the most important
questions we must confront when faced with a situation involving the
gravitational field is: on what background spacetime should one
consider the effects of quantum fluctuations? In the case of
gravitational collapse, as considered in this paper, we have proposed a
background metric radically different from those normally considered. Our
motivation for doing this stemmed from the not unreasonable requirement
that gravitational collapse should actually take place via a unitary
evolution. Given that there is an unambiguous metric at very early
times, i.e. one which describes a collapsing star plus other possible
quantum fields, which classically would form a black hole, we argued
that a metric at very late times compatible with unitary evolution would be
that of a white hole describing the efflux of matter. We subsequently
argued that the two metrics could be made globally compatible with each
other in a regular way, modulo a possible conical singularity just
outside the putative horizon of the collapsing star. Hence the
requirement of quantum coherence led us to ``derive'' a background
metric very different from that of a ``normal'' black hole. We claim
that this metric would be a solution of the classical Einstein
equations everywhere except in the vicinity of the conical
singularity/bounce.

Given this globally defined background metric one would wish to study
small deviations, $\d_{\rm in}$ and $\d_{\rm out}$, in the in- and
out-states, which are defined with respect to this metric. If the
deviations are not small then one would find that the c-number metric
previously found would not be suitable.  In hard and soft particle
language what one requires is a background metric derived from the
``hard particle Einstein equations'' which when the back reaction due
to the soft particles is introduced does not change very much.  In the
locality of the conical singularity it would be quite difficult to give
an unambiguous definition of the metric due to the difficulty in
distinguishing between matter and gravity in such a region. A priori
the correct metric to choose here is not exactly known, because we are
potentially dealing with a region where high concentrations of
particles occur with relative energies beyond the Planck regime. In
many cases though one can guess. In particular if the region with
unknown interactions is concentrated at a point (in the space of
longitudinal coordinates) then the most reasonable thing one can choose
is that the metric is continuous there, and that where few particles
occur it will be flat. This is all our bounce model amounts to.

One can take these notions one step further. Consider a given hard
particle background. We advocated that it has to be topologically
trivial, and moreover, all soft particle configurations will generate
only topologically trivial metrics. Considered in a functional integral
setting we would then maintain that for gravitational collapse one
should sum over only those soft particle configurations appropriate to
a particular topologically trivial hard particle background. One could
then consider another hard particle background, e.g. one with a
different mass; however, we would maintain that it also should be
topologically trivial. If one sums over different possible hard
particle backgrounds, with appropriate weights one should be able to
obtain the full quantum gravity functional integral. But this
immediately leads to an interesting proposal for an unambiguous
prescription for the entire functional integral.
In our proposal one can proceed by:
\smallskip
\noindent{\it Integrating only over topologically trivial Lorentzian
metrics to get the complete amplitude}.
\smallskip
\noindent If one accepted this proposal then space-times with
multiple universes, wormholes
and all other such concoctions would be excluded from the functional
integrand.

Note that the structure of ${\cal I}^+$ is pretty much the same both in
our bounce spacetime (Fig.~3), or in that of an evaporating black hole
(Fig.~1). The two spacetimes differ greatly, however, behind the horizon,
or bounce. One of the main objections to unitarity in gravitational
collapse has been the apparent acausality associated with trying to
retrieve information from behind the horizon if unitarity is to be
manifest on ${\cal I}^+$. In theories with remnants unitarity would not
be manifest in the Hawking radiation alone but requires correlations
with the remnant also to be accounted for. In our bounce model
unitarity is manifest (at least in the soft particle sector) on ${\cal
I}^+$. The Hawking radiation itself is globally pure. Information did
not have to be expunged from the ingoing matter, nor was there any
acausal retrieval of information, all because in our bounce model there
was no event horizon. All the field degrees of freedom associated with
$r<\rbn$ we can count as ``internal'' degrees of freedom associated
with our black hole ``doppelganger''. Their total entropy will be given
by the usual area law. In stark distinction to the case of a true black
hole, however, these internal degrees of freedom can be reconstituted
on ${\cal I}^+$. This can clearly be seen in Fig.~3 where there is
nothing to prevent information getting to ${\cal I}^+$, after the
bounce or stop, along ${\rm F}\iota^+$. This is a completely causal
process.

One might also think of the
quantum mechanical degrees of freedom as
being ``duplicated'' in the Hawking radiation. This duplication is not
in conflict with the quantum mechanical superposition principle if one
insists that `superobservers' are an impossibility. A superobserver is
a detector that measures simultaneously objects that went across the
horizon, and Hawking radiation emerging at later times. We argued
earlier that the divergent commutators of the observables measuring
Hawking radiation on the one hand and the observables inside the
horizon on the other prevent their simultaneous measurements.

A simple model of the duplication process is the following. Consider a
particle described by a Hamiltonian $H_1(t)$ , with
$-\infty<t<+\infty$.  Suppose now that there are two ways to describe
its quantum states. One is by simply specifying $\j_1(t)$ as being the
solution of the Schr\"odinger equation with $H=H_1(t)$ . The other is
by taking $\j_2(t) =\j_1(t)$ for $t\leq a$, but ${\rm
d\j_2/d}t=-iH_2(t)\j_2$ during $a<t<b$, with $H_2 \ne H_1$ , and again
the original Schr\"odinger equation for $t>b$.  We are talking about
the same system, and the same state, but at $t\gg b$ we seem to have
two wave functions, $\j_1$ and $\j_2$. This may resemble the situation
for black holes.  $\j_1$ corresponds to the wave functions as seen by
an outside observer and $\j_2$ to what is seen by observers who crossed
the horizon. Of course one can also say that since $H_1\ne H_2$ one
simply has a violation of general relativity; the physical system seen
by the infalling observer no longer corresponds with what is seen
outside. Because it will be forever impossible for the two observers
later to compare their data there will never be any conflict.

We believe that, in principle at least, our methodology allows for any
calculation to be done. In practice however there are many limitations,
not the least of which is the tendency for ingoing matter in the
background of our preferred space-time metric to generate very large
blueshifts thereby bringing into question the validity of one's metric
as a good background around which to expand. Even mild modifications of
the wavefronts of ingoing particles can cause severely energetic
showers of particles to come out. The escalating gravitational field of
the collapsing star can turn soft ``in'' particles into very hard
``out'' particles or vice versa. In our explicit calculation of section
6 we circumvented this problem by having a bounce at a radius
sufficiently far from the horizon that extremely large blue shifts were
not generated, but still close enough that one could generate
asymptotic Hawking radiation with an approximately Planckian spectrum.
In this calculation we would also run into another problem with hard
particles, not due to the fact that individual particles can become
hard, but that the radiation of soft particles over a sufficiently long
time would remove enough energy from the collapsing star that the back
reaction due to the emitted soft particles may not be neglected. The
back reaction in this case however may perhaps be easier to deal with
than when individual soft particles become hard.

It is evident that many of the difficulties one encounters in
gravitational collapse, certainly in the approach we have taken here,
are due to large back reaction effects. In the past many authors have
argued that the back reaction due to Hawking radiation is small, and
therefore the classical metric of a collapsing star will proffer a good
background around which to expand. We believe this not to be true. We
have argued in this paper that the back reaction is large. For us, the
bounce model gives additional reason for believing it to be large. What
we mean by ``large'' and ``small'' back reaction however, depends
on choice of coordinate frame as well as the states considered. An
important feature of the bounce model is the very great apparent
asymmetry between retarded and advanced time . The shell radiates
energy only between the collapse time, $\tin$, and the bounce time
$\tbn$. The emitted radiation, as seen in the coordinates used in
section 6, thus occupies a span of retarded time which can be very
long, and a span of advanced time which can be very short. Thus the
back reaction in retarded time and advanced time will appear very
different. Note that the above asymmetry is not of a
``fundamental'' kind, as we have emphasized several times our formalism
is manifestly time reversal invariant, but arises due to our freedom to
choose asymmetric looking coordinate frames, i.e. to go to coordinate
frames where soft particles can look hard and vice versa. The task of
computing the back reaction in a model such as the bounce model is an
important one to which we will return in another paper.

Our claim is that the bounce metric may offer a more self-consistent
background field around which to calculate quantum corrections. As
opposed to postulating a background metric one would of course prefer
to derive it from the full quantum equations of motion. Such a
calculation at the present time seems beyond us. Ultimately of course
we do not believe that the bounce can be truly singular but must be
smeared. We have taken it to occur outside the $R=2M$ surface of the
star. However, as the star collapses it Hawking radiates.  After a
certain amount of time, wherein the star has a new mass $M'$, the new
effective Schwarzschild radius of the star will be $2M'$. The bounce
could now be taken to occur just outside $2M'$ rather than $2M$. The
precise nature of the bounce will depend strongly on whether one can
regard the outgoing Hawking radiation as hard or soft. If one can think
of it as being soft then it may be possible to push back the bounce to
the point at which the mass of the star is such that the energy density
in the Hawking radiation is very large. The bounce would then represent
the final catastrophic explosion of the black hole. Alternatively, if
we thought of the outcoming radiation as being hard, then we can think
of a series of bounces terminating in a final one at the Planck scale.
In both these scenarios what we get in place of a ``remnant'' is a
violent explosion of particles with Planckian energies. This will
always be at the edge of what we can handle with standard quantum
theories.

We have not been able to show that the procedure advocated by us will
lead to a completely unitary $S$-matrix, in terms of both hard and soft
particles. What we instead have argued is that we have access to the
$S$-matrix for soft particles on a topologically trivial hard particle
background, and that this $S$-matrix is indeed unitary if the conical
singularity is sufficiently smeared.  A different topologically trivial
hard particle background would also lead to an unitary $S$-matrix.
Since the unitary $S$-matrices we obtained this way only span part of
the entire Hilbert space, it is not obvious that the complete
$S$-matrix obtained by combining all of them will be unitary.  We do
believe however that we have a promising avenue of investigation for
the future with which to examine this question.
\bigskip
\noindent{\bf ACKNOWLEDGEMENTS}
\medskip
CRS is grateful to FOM for financial support. BFW acknowledges both FOM
and NWO for support.
\bigskip
\line{\bf REFERENCES \hfil}

\vglue 0.4cm \item{1.}S.W.~Hawking, {\it Commun. Math. Phys.} {\bf 43}
(1975) 199; W.G. Unruh, Phys. Rev. D14 (1976) 870; R.M.~Wald, {\it
Commun. Math. Phys.} {\bf 45} (1975), 9; J.B.~Hartle and S.W.~Hawking,
{\it Phys.  Rev.} {\bf D13} (1976) 2188; S.W.~Hawking, {\it Phys. Rev.}
{\bf D14} (1976) 2460; S.W. Hawking, {\it Commun. Math. Phys.} {\bf 87}
(1982) 395.

\item{2.} T.~Banks, L.~Susskind and M.E.~Peskin, {\it Nucl. Phys.} {\bf
B244} (1984) 125; J.~Ellis, J.~Hagelin, D.V.~Nanopoulos and
M.~Srednicki, {\it Nucl.  Phys.} {\bf B241} (1984) 381.

\item{3.} C.G.~Callen, S.B.~Giddings, J.A.~Harvey and A.~Strominger,
{\it Phys. Rev.} {\bf D45} (1992) R1005.

\item{4.} G. 't Hooft, {\it Nucl. Phys.} {\bf B335} (1990) 138.

\item{5.} G. 't Hooft, {\it Nucl.Phys.} {\bf B256} (1985) 727.

\item{6.} L.~Susskind, L.~Thoraclius and J.~Uglum, {\it ``The Stretched
Horizon and Black Hole Complementarity''}, Stanford preprint
SU-ITP-93-15.

\item{7.} J.D.~Bekenstein, {\it Phys. Rev.} {\bf D5} (1972) 1239.

\item{8.} G.~'t Hooft, "On the Quantization of Space and Time", Proc.
of  the  4th Seminar on Quantum Gravity, May 25-29, 1987, Moscow, USSR,
ed. M.A.~Markov et al (World Scientific 1988).

\item{9.} G.~'t Hooft, {\it in} Proceedings of the Int. Conf. on
Fundamental  Aspects of Quantum Theory, to celebrate the 60th birthday
of Yakir Aharonov, December 10-12, 1992, Columbia, SC., to be pub.

\item{10.} J.B.~Hartle, {\it in} Quantum Gravity 2, ed.s C.J.~Isham,
R.~Penrose and D.W.~Sciama (Clarendon Press 1981).

\item{11.} N.D.~Birrell and P.C.W.~Davies, {\it Quantum Fields in
Curved Space} (CUP 1982).

\item{12.} P.C.W.~Davies, S.A.~Fulling and W.G.~Unruh, {\it Phys. Rev.}
{\bf D13} (1976) 2720.

\end